\documentclass[a4paper, 9pt]{article}
\pdfoutput=1

\usepackage{INTERSPEECH2022}
\usepackage{siunitx}
\usepackage{url}
\usepackage{cleveref}
\usepackage{tikz}
\usetikzlibrary{positioning, shapes, calc, arrows} 
\usepackage{placeins}
\usepackage{glossaries}
\usepackage{subcaption}
\usepackage{adjustbox}
\usepackage{xcolor}
\usepackage{import}
\usepackage{pgfplots}
\usepgfplotslibrary[groupplots]
\pgfplotsset{compat=newest}

\newacronym{vtlp}{VTLP}{Vocal Tract Length Perturbation } 
\newacronym{in}{IN}{Instance Normalization}
\newacronym{vad}{VAD}{Voice Activity Dection}
\newacronym{VC}{VC}{Voice Conversion}
\newacronym{jnd}{JND}{Just Noticeable Difference}
\newacronym{FVAE}{FVAE}{Factorized Variational Autoencoder}
\newacronym{CPC}{CPC}{Contrastive Predictive Coding}
\newacronym{GAN}{GAN}{Generative Adversarial Network}
\newacronym{eer}{EER}{Equal Error Rate}
\newacronym{wer}{WER}{Word Error Rate}
\newacronym{mse}{MSE}{Mean-Squared Error}

\title{
    Investigation into Target Speaking Rate Adaptation for Voice Conversion
}
\name{
    Michael Kuhlmann$^1$, Fritz Seebauer$^2$, Janek Ebbers$^1$,
    Petra Wagner$^2$, Reinhold Haeb-Umbach$^1$
}
\address{
  $^1$Paderborn University, Germany\qquad
  $^2$Bielefeld University, Germany
}
\email{
    \{kuhlmann,ebbers,haeb\}@nt.uni-paderborn.de,\\
    \{fritz.seebauer,petra.wagner\}@uni-bielefeld.de
}

\begin{document}

\maketitle
\begin{abstract}
    Disentangling speaker and content attributes of a speech signal into separate latent representations followed by decoding the content with an exchanged speaker representation is a popular approach for voice conversion, which can be trained with non-parallel and unlabeled speech data.
    However, previous approaches perform disentanglement only implicitly via some sort of information bottleneck or normalization, where it is usually hard to find a good trade-off between voice conversion and content reconstruction.
    Further, previous works usually do not consider an adaptation of the speaking rate to the target speaker 
    or they put some major restrictions to the data or use case.
    Therefore, the contribution of this work is two-fold.
    First, we employ an explicit and fully unsupervised disentanglement approach, which has previously only been used for representation learning, and show that it allows to obtain both superior voice conversion and content reconstruction.
    Second, we investigate simple and generic approaches to linearly adapt the length of a speech signal, and hence the speaking rate, to a target speaker and show that the proposed adaptation allows to increase the speaking rate similarity with respect to the target speaker.

\end{abstract}

\section{Introduction}
\gls{VC} deals with the transfer of speaker characteristics from one utterance to another while keeping the content information untouched.
While some works \cite{sisman2020overview} learn the mapping from one speaker to another using parallel data, i.e., data which contains same sentences uttered by different speakers, another course of research focuses on learning \gls{VC} from non-parallel data, i.e., where each sentence is only uttered by a single speaker.
Different approaches exist for non-parallel \gls{VC} learning, e.g., based on \glspl{GAN} \cite{kaneko2019stargan} or based on information perturbation \cite{choi2021neural}.
Another particularly popular approach to non-parallel \gls{VC} is to perform disentanglement via autoencoding.
The idea is to encode content and speaker information into two disjoint representations by, e.g., using a tight bottleneck \cite{Qian2019, Wu2020b}, normalization layers \cite{Wu2020b, Chou2019, Chen2020} or an adversarial approach \cite{Chou2018}.
Then, at test-time, \gls{VC} can be performed by decoding the content representation of an utterance together with an exchanged speaker representation.
So-called \emph{zero}- \cite{Qian2019} or \emph{one-shot} \cite{Wu2020b, Chou2019, Chen2020} approaches can perform any-to-any voice conversion, where neither source nor target speakers need to be in the training data.

Most of above works mainly tackle timbre conversion, where the length of the source utterance, and hence the speaking rate, is kept unchanged.
However, speaking rate is also an important characteristic of individuals \cite{smith1975effects}
or situation specific speaking styles~\cite{lindblom1990explaining}.
There are some works 
which jointly learn to also convert the speaking rate using sequence-to-sequence~\cite{tanaka2019atts2s, hayashi2021non, zhang2019non} or disentanglement~\cite{qian2021global, wang2021adversarially} approaches.
These works, however, require parallel data during training or known speaker identities during test time.
In \cite{prananta2022effectiveness} it was shown that speech recognition for dysarthric speech can be enhanced by a simple linear interpolation of an utterance's spectrogram and outperforms cycle-based \gls{VC} approaches on parallel data.

The contribution of this paper is two-fold.
First, we propose to perform \gls{VC} with our fully unsupervised \gls{FVAE} model \cite{Ebbers2020}.
The \gls{FVAE} model, which achieves disentanglement by an adversarial \gls{CPC} loss, has, so far, only been considered for learning disentangled features.
We here investigate its suitability for \gls{VC} when decoding the disentangled content representation with an exchanged speaker representation.
Second, we investigate ways to estimate an interpolation factor for linearly adapting the speaking rate to the target speaker.
We consider two variants.
The first is based on an estimation of the phoneme rate of an utterance from transcribed training data, which then runs fully unsupervised at test-time.
The second variant is based on acoustic unit rates from an unsupervised segmentation model and neither needs text transcriptions at train nor at test time.
With our proposed fully unsupervised \gls{VC} model combined with the fully unsupervised speaking rate adaptation, our model can be trained on and applied to fully unsupervised raw speech signals.
We compare our approach against AutoPST~\cite{qian2021global}, a state-of-the-art model for voice and prosody conversion without text transcriptions or parallel data.
We show that a simple post-processing for speaking rate adaptation can achieve the desired effect of increased target speaker similarity and outperforms AutoPST in voice and speaking rate similarity.

\vspace{-2mm}
\section{Target speaker conversion}

\begin{figure}[h]
    \centering
    \resizebox{0.45\textwidth}{!}{
        \begin{tikzpicture}[
            >=stealth, semithick, auto, scale=1,
            box/.style={rectangle, draw, align=center, minimum width=2cm, minimum height=1cm, rounded corners},
        ]
            \node[minimum width=1.5cm] (x_content) {$\mathbf{X}_\text{content}$};
            
            \node[box] at ($(x_content) + (2.8, -0.25)$) (vc) {VC Model}; 
            \node[box, right=0.7cm of vc, minimum width=0.5cm] (vocoder) {Vocoder}; 
            
            \node[below=1cm of x_content, minimum width=1.5cm] (x_style) {$\mathbf{X}_\text{style}$};
            
            \node[box, right=1cm of x_style] (sr_estimator_style) {Speaking Rate\\Estimation};
            \node[box, below=0.2cm of sr_estimator_style] (sr_estimator_content) {Speaking Rate\\Estimation};
            
            \node[draw, rectangle, right=0.7cm of vocoder, minimum height=1cm] (wsola) {WSOLA};
            
            \node[right=0.5cm of wsola] (y_vc) {$\tilde{y}_\text{VC}$};
            
            \node[right=3.08cm of sr_estimator_style, minimum width=0.5cm, inner sep=0pt, circle, draw, ] (mul) {};
            \path [{}-{}, draw]
                (mul.north east) -- (mul.south west)
                (mul.north west) -- (mul.south east);
            \node[right=2.83cm of sr_estimator_content, minimum width=1cm, draw, rectangle] (div) {$\frac{1}{\text{x}}$};
            
            \path [{}-{>}, draw]
                (x_style) -- ($(x_style) + (1.25, 0)$) node[]{} -- +(0, 1) node[]{} -- ($(vc) + (-1.02, -0.26)$) node[]{};
            
            \path [{}-{>}, draw]
                ($(x_style) + (1.25, 0)$) node[circle, draw, fill=black, inner sep=0pt, minimum size=2pt]{} -- (sr_estimator_style);
            
            \path [{}-{>}, draw]
                (x_content) edge[] node[]{} ($(vc) + (-1.02, 0.25)$)
                ($(x_content) + (1, 0)$) node[circle, draw, fill=black, inner sep=0pt, minimum size=2pt]{} -- +(0, -2.74) -- (sr_estimator_content)
                (vc) edge[] node[midway, above]{$\mathbf{X}_\text{VC}$} (vocoder)
                (sr_estimator_style) edge[] node[]{} (mul)
                (sr_estimator_content) edge[] node[]{} (div)
                (div) edge[] node[]{} (mul)
                (mul) edge[] node[]{} (wsola)
                (vocoder) edge[] node[midway, above]{$y_\text{VC}$} (wsola)
                (wsola) edge[] node[]{} (y_vc);
        \end{tikzpicture}
    }
    \caption{Overview of the proposed approach}
    \label{fig:overview_approach}
\end{figure}
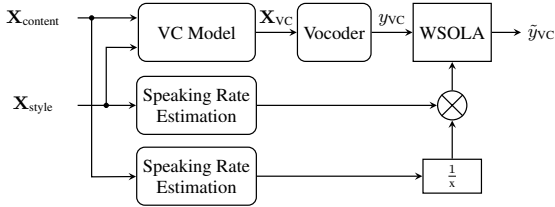

Given an utterance from an arbitrary speaker we want to match the voice and speaking rate to a target speaker while keeping the content unchanged.
We want to impose as few restrictions as possible on the system.
Specifically, the conversion process should work for arbitrary target speakers, needs only one utterance from the target speaker and does not rely on any text transcriptions.
For this, we consider two successive steps:
\begin{enumerate}
    \item Any-to-any \gls{VC}:
    We use a neural-based disentanglement system that separates content from speaker style and swap in the target speaker style to perform a voice conversion.
    The output is a Mel spectrogram and we use a pretrained neural vocoder to obtain the converted audio signal.
    \item Speech rate conversion:
    We use WSOLA~\cite{verhelst1993overlap} to globally adapt the tempo of the converted signal to match the speaking rate of the target utterance. The interpolation factor is the ratio of target phoneme rate to source phoneme rate. To obtain the phoneme rates from the utterances, we investigate two approaches.
\end{enumerate}
With this two-step approach, we can decouple the speaking rate conversion from the \gls{VC} model and the vocoder to separately evaluate the influence of adding the speaking rate conversion to the voice conversions.
It is also possible to integrate the speaking rate conversion into the vocoder, e.g., non-trainable analysis-synthesis vocoders like STRAIGHT and WORLD or into non-autoregressive \gls{VC} models~\cite{lee2022duration}.
We leave an analysis of the benefits and drawbacks of each of these approaches to future work.
~\Cref{fig:overview_approach} summarizes the proposed approach.

\vspace{-0.4cm}
\subsection{Voice conversion model}
\label{sec:vc-model}
\vspace{-2mm}
To perform the voice conversion, we use an autoencoder to disentangle speaker information from the content and pair the content embedding with a neural speaker embedding from the target speaker.
We use the \glsfirst{FVAE} with an adversarial contrastive loss on the content embedding which achieves a well-balanced trade-off between informative linguistic content and speaker invariance~\cite{Ebbers2020}.

\begin{figure}[h]
    \centering
    \resizebox{0.45\textwidth}{!}{
        \begin{tikzpicture}[
            >=stealth, semithick, auto, scale=1,
            box/.style={rectangle, draw, align=center, minimum width=2cm, minimum height=1cm, rounded corners},
        ]
            \node[box] (style-encoder) {Style\\Encoder};
            
            \node[left=1cm of style-encoder, minimum width=1.5cm] (x) {$\mathbf{X}$};
            
            \node[rectangle, draw, below=0.5cm of x] (vtlp) {VTLP};
            
            \node[rectangle, draw, below=0.5cm of vtlp, align=center, minimum width=1.5cm]
            (in) {Instance\\Norm};

            \node[box, right=1cm of in] (z-encoder) {Content\\Encoder};
            
            \node[draw, rectangle, right=0.5cm of style-encoder] (gap) {GAP};
            \coordinate (gap brc) at (gap.south east);
            
            \node[box, right=0.5cm of gap brc] (decoder) {Decoder};
            
            \node[right=0.5cm of decoder] (xhat) {$\mathbf{\hat{X}}$};
            \coordinate (decoder tlc) at (decoder.north west);
            \coordinate (decoder blc) at (decoder.south west);
        	\path let
        		\p{A} = (gap),
        		\p{B} = (decoder tlc)
        		in
	            coordinate (s-in) at ( \x{B}, \y{A});
	       \path let
	            \p{A} = (z-encoder),
	            \p{B} = (decoder blc)
	            in
	            coordinate (z-dec) at (\x{B}, \y{A});
	       
	       \node[draw, rectangle] at ($(z-dec) + (-1., 1)$) (sample) {Sample};
	       
	       
	       
	       \node[box, right=1.85cm of z-encoder] (fcpc) {$f_\text{cpc}$};
	       
	       \node[right=0.5cm of fcpc] (h) {$\mathbf{H}$};
	       
	       \node[minimum size=2pt, fill=black, inner sep=0pt, circle, draw,] at ($(z-dec) + (-1., 0)$) () {};
            
            \path [{}-{>}, draw]
                (x) edge[] node[]{} (vtlp)
                (x) edge[] node[]{} (style-encoder)
                (vtlp) edge[] node[]{} (in)
                (in) edge[] node[]{} (z-encoder)
                (style-encoder) edge[] node[above, midway]{$\mathbf{S}$} (gap)
                (gap) edge[] node[above, midway]{$\mathbf{\bar{s}}$} (s-in)
                (decoder) edge[] node[]{} (xhat)
                ($(z-dec) + (-1.5, 0)$) -- (fcpc);
            \path [{}-{>}, draw]
                (z-encoder) -- ($(z-dec) + (-1., 0)$) node[midway, below]{$\mathbf{\tilde{Z}}$} -- (sample);
            \path [{}-{>}, draw]
                (sample) -- ($(decoder blc) + (-1., 0.3)$) -- ($(decoder blc) + (0, 0.3)$) node[below, midway]{$\mathbf{Z}$};
            \path [{}-{>}, draw]
                (fcpc) edge[] node[]{} (h);
        \end{tikzpicture}
    }
    \caption{The architecture of the FVAE \cite{Ebbers2020}}
    \label{fig:fvae}
    \vspace{-2mm}
\end{figure}
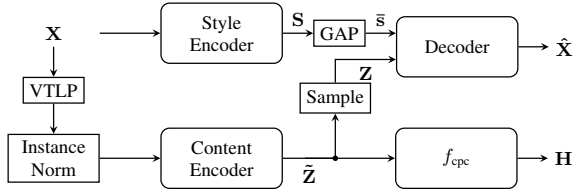

The FVAE consists of two fully-convolutional encoders and a fully-convolutional decoder (see Figure~\ref{fig:fvae}).
The style encoder extracts a sequence of style embeddings $\mathbf{S}$ from the input feature map $\mathbf{X}$.
Static information is aggregated into a single embedding $\mathbf{\bar{s}}\in\mathbb{R}^{D_s}$ using global average pooling (GAP) over time on $\mathbf{S}$.
Simultaneously, the content encoder extracts a content representation $\mathbf{Z}$ from the input.
To drive the decoder to infer the timbre from $\mathbf{\bar{s}}$ rather than from $\mathbf{Z}$ during reconstruction, $\mathbf{X}$ is first distorted by \gls{vtlp}~\cite{jaitly2013vocal} and then transformed by \gls{in}~\cite{ulyanov2016instance} before being passed to the content encoder.
For each time frame, the content encoder outputs a tuple $\mathbf{\tilde{z}}_n=(\boldsymbol{\mu}_{z_n}, \boldsymbol{\xi}_{z_n})$, which are mean and log-variance vectors, respectively, and $\mathbf{Z}$ is obtained through the reparameterization trick \cite{Kingma2014a} during training.
Then, $\mathbf{\bar{s}}$ is frame-wise concatenated with $\mathbf{Z}$ and passed to the decoder, yielding the reconstructed speech feature $\mathbf{\hat{X}}$.
Both encoders and the decoder are jointly trained to minimize the frame-wise \gls{mse} between $\mathbf{X}$ and $\mathbf{\hat{X}}$.

To foster disentanglement, a contrastive loss~\cite{Oord2018} is applied on the content embedding.
An adversarial encoder $f_\text{cpc}$ extracts an embedding {$\mathbf{h}_n=f_\text{cpc}(\mathbf{\tilde{z}}_n)$} from the content encoder output for each time frame $n$.
A prediction head $g_k(\cdot)$ is used to yield a prediction $\mathbf{\hat{h}}_n=g_k(\mathbf{h}_{n-k})$ for a future time step in the embedding dimension, where $k$ is the lookahead shift. 
The contrastive objective then tries to find the correct prediction out of a number of candidates.
To remove static information from $\mathbf{\tilde{Z}}$, a large lookahead shift $k$ is used (corresponding to \SI{1}{s}) and the content encoder is trained to maximize the contrastive loss.
We use the same model architecture and training procedure as in~\cite{Ebbers2020} except that we set the size of the latent content embedding to $D_\mathbf{z}=16$ and use a downsampling factor of $S_\text{ds}=2$.

Given the utterances $\mathbf{X}_\text{content}$ and $\mathbf{X}_\text{style}$, from which the content and style is to be extracted, respectively, we extract $\boldsymbol{\mu}_\mathbf{Z}$ from $\mathbf{X}_\text{content}$ and $\mathbf{\bar{s}}$ from $\mathbf{X}_\text{style}$ which are passed to the decoder to yield the converted speech feature $\mathbf{X}_\text{VC}$.
We use Mel spectrograms to represent the speech signals.
Waveform synthesis is performed with a pretrained
ParallelWaveGAN~\cite{Yamamoto2020}.

\vspace{-2mm}
\subsection{Speaking rate estimation}
\vspace{-2mm}
To represent the speaking rate of an utterance, we opted for the utterance phoneme rate which is the number of phonemes in the utterance $C_p$ over the total utterance duration $N$ in seconds or samples, excluding silence segments.
Another viable option could be the syllable rate~\cite{pfitzinger2002reducing, jiao2016online}.
We use the phoneme rate due to the availability of corpora with hand-annotated phoneme labels or their likewise reliable estimation from transcriptions using forced alignment tools which allows us compare the results to an unsupervised speaking rate estimator~(\Cref{sec:scenario-2}).
The utterance duration can be simply measured after applying a \gls{vad}.
To build a predictor for the number of phonemes, we consider two scenarios.

\vspace{-2mm}
\subsubsection{Scenario 1: Text-annotated training data}
\vspace{-2mm}
If transcribed audio data is available, we can perform a forced alignment to obtain the phoneme annotations.
However, as mentioned above we do not want to be dependent of any transcriptions during inference time.
Using the phoneme alignment, we train a local phoneme rate predictor which captures segmental variations in the speaking rate.
We then compute the number of phonemes in the utterance by integrating over the local phoneme rate.
This way, we can estimate the number of phonemes in the utterance without having the text transcription at inference time.

Following the definition from~\cite{pfitzinger2002reducing}, the local phoneme rate $r_p(n)$ is the smoothed average of the instantaneous phoneme rate ${I_p(n),\, 0\leq{}n\leq{}N-1}$, with a window function $w(n)$:
\begin{equation}
    r_p(n) = \frac{
        \sum_{m=0}^{K-1} I_p\left(n+m-\lfloor\frac{K-1}{2}\rfloor\right)w(m)
    }{\sum_{m=0}^{K-1} w(m)}, 
    \label{eq:lpr}
\end{equation}
where $w(n)$ is a Hann window of size $K$,  and $I_p(n)$ is zero-padded at both sides.
The instantaneous phoneme rate is a step-wise function where the step widths are given by the phoneme duration and the step heights are given by the inverse duration
\begin{equation}
    I_p(n) = \frac{1}{S_{i+1}-S_i}, \quad S_i\leq{}n\leq{}S_{i+1},
\end{equation}
where $S_i$ and $S_{i+1}$ denote the start and stop times of the $i$-th phoneme.
From this definition, it follows that ${\int_{-\infty}^\infty I_p(n)dn=C_p}$.

We train a local phoneme rate predictor consisting of stacked dilated 1d-convolution layers, two LSTM layers and a final convolution layer with softplus activation.
The receptive field of the CNN is chosen to match the window size of $w(n)$ which is set to \SI{625}{ms}~\cite{pfitzinger2002reducing}.
We use Mel spectrograms with delta and delta-delta features as input and minimize the \gls{mse} between the prediction and true local phoneme rate.
During inference, the number of phonemes can then be retrieved from the predicted local phoneme rate.

\vspace{-2mm}
\subsubsection{Scenario 2: Unlabeled training data}
\label{sec:scenario-2}
\vspace{-2mm}
In the case that we do not have any annotated speech audio available, we can cast the estimation of the number of phonemes into an unsupervised phoneme segmentation task and count the number of predicted boundaries.
Several works tackle phoneme segmentation without labels of which we chose one particularly simple approach that showed to generalize well to out-of-domain data~\cite{kreuk2020self}.

A strided CNN takes the raw waveform as input and encodes it into a latent sequence vector.
The encoder is trained to predict the next frame from the current one using a contrastive objective~\cite{Oord2018}.
To predict the phoneme boundaries from the latent sequence vector, a score function is used which measures how dissimilar two adjacent frames are:
High dissimilarity indicates a phoneme change.
A peak detection algorithm is run over the dissimilarity curve and the estimated peaks are the predictions for the phoneme boundaries.
We use the open-source implementation from~\cite{kreuk2020self} to train the model on the Timit+ training set.

\vspace{-3mm}
\section{Experiments}
We evaluate the voice and speaking rate conversion approach on the CSTR VCTK corpus~\cite{Yamagishi2019} 
which contains utterances from 110 English speakers.
We split the VCTK dataset into three subsets for training, validation and testing with respectively 70, 20 and 20 speakers per subset.
During training, we only use 90\% of the utterances in the training and validation subsets and leave the remaining 10\% for a closed speaker set evaluation.
To train the local phoneme rate predictor, we trim the beginning and end silence and discard utterances which are shorter than \SI{2}{s} after trimming.
The forced alignments for scenario 1 training and speaking rate conversion evaluation are obtained with the Montreal Forced Aligner~\cite{mcauliffe2017montreal}.

\vspace{-2mm}
\subsection{Accuracy of utterance phoneme rate prediction}
\vspace{-2mm}
We first evaluate how well we can predict the utterance phoneme rate from the speech signal under scenario 1 and 2.
Starting off with scenario 1, we evaluate the local phoneme rate predictor in terms of the Pearson correlation coefficient $r$ to the reference local phoneme rate.
We compute the number of phonemes from the predicted local phoneme rate and evaluate the absolute difference $e_{C_p}$ to the number of phonemes which were annotated from the forced alignment.
We also show the relative error $e_{C_p}/C_p$ which is equal to the relative difference in the utterance phoneme rates between prediction and target (the utterance duration $N$ is the same for both).
\Cref{tab:eval-sc1} shows the results for the local phoneme rate predictor.
The predicted utterance phoneme rate is a reliable substitute for the true one
because their relative difference is below the perception threshold of 5\%, i.e., human listeners will likely not perceive any difference in tempo when listening to utterances with these phoneme rates given that the content is the same (see \Cref{sec:conversion}).

\begin{table}[b]
    \centering
    \caption{
        Performance of the local phoneme rate predictor evaluated against the ground truth local phoneme rate which is computed from the forced alignment. 
    }
    \label{tab:eval-sc1}
    \vspace{-3mm}
    \begin{tabular*}{0.45\textwidth}{@{\extracolsep{\fill}}c c c c}
        \toprule[1.5pt]
        Test set & $r$~$\uparrow$ & $e_{C_p}~\downarrow$ & ${e_{C_p}}/{C_p}~\downarrow$ \\
        \midrule
        Seen speakers & $.8319$ & $2.29$ & $4.14\%$ \\ 
        Unseen speakers & $.8064$ & $2.60$ & $4.62\%$ \\ 
        \bottomrule[1.5pt]
    \end{tabular*}
\end{table}

\begin{table}[t]
    \centering
    \caption{
        Quality of unsupervised phoneme segmentation~\cite{kreuk2020self}.
        Precision (P) and Recall (R) are calculated with a \SI{20}{ms} tolerance window.
        All values are given in percent.
    }
    \label{tab:eval-sc2}
    \vspace{-3mm}
        \begin{tabular*}{0.45\textwidth}{@{\extracolsep{\fill}}c c c c c c}
            \toprule[1.5pt]
            Test set & P $\uparrow$ & R $\uparrow$ & F1 $\uparrow$ & R-Val $\uparrow$ & ${e_{C_p}}/{C_p}~\downarrow$ \\
            \midrule
            Timit & 82.93 & 83.04 & 82.99 & 85.48 & 11.50 \\ 
            VCTK & 69.70 & 72.34 & 71.00 & 74.92 & 11.43 \\
            \bottomrule[1.5pt]
        \end{tabular*}
    \vspace{-1mm}
\end{table}

\Cref{tab:eval-sc2} shows the evaluation of the unsupervised segmentation on matched (Timit) and unmatched (VCTK unseen speakers) test sets.
While the results for the matched case coincide with~\cite{kreuk2020self} there is a noticeable performance drop for the unmatched case caused by the domain mismatch.
Still, the relative difference in the utterance phoneme rate measured by $e_{C_p}/C_p$ is similar for both test sets.
While Timit has a higher ratio of predicted boundaries within the tolerance window (avg. $29.70$ of $37.48$) than VCTK (avg. $44.78$ of $65.24$) the total number of predicted boundaries is surprisingly accurate for VCTK (avg. $61.34$ phonemes / sentence).
Note also that the segmentation is applied to both the source and the target speaker's utterance.
Thus,  the absolute segmentation quality may not be that important, as long as similar errors are made in both utterances.

\vspace{-2mm}
\subsection{Voice and speaking rate conversion}\label{sec:conversion}
\vspace{-2mm}

\begin{figure}[b]
    \centering
    \begin{subfigure}[h]{0.45\textwidth}
        \includegraphics[width=0.95\textwidth]{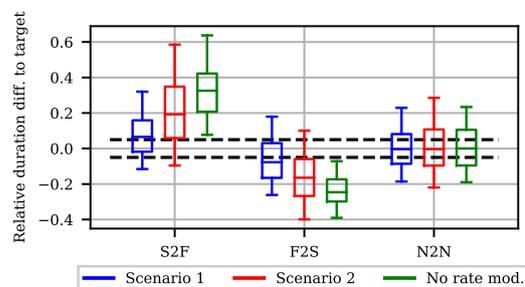}
    \end{subfigure}
    \vspace{-3mm}
    \caption{Normalized duration difference after speaking rate conversion on the first 24 parallel sentences in VCTK with unseen speakers.
    The black dotted lines mark the \glsfirst{jnd} for speech tempo.
    }
    \label{fig:duration-diff-vctk}
\end{figure}

In order to measure whether the proposed approaches can effectively perform a speaking rate conversion, we rank the speakers in the test set by increasing average phoneme rate and perform a 3-fold split.
We denote these splits as slow, normal and fast, respectively.
The first 24 sentences in the VCTK corpus are parallel, i.e., they are read by all speakers.
As an objective measure for the speaking rate conversion we evaluate the difference between the utterance duration after conversion and the duration of the parallel target sentence normalized to the target duration.
The utterance phoneme rate for the target speaker is extracted from a different sentence from the target speaker.
We compare the results under scenario 1 and 2 to the case when no speaking rate modification is performed.
\Cref{fig:duration-diff-vctk} shows the results for the slow-to-fast (S2F), fast-to-slow (F2S) and normal-to-normal (N2N) conversion.
The black dotted lines mark the \gls{jnd} which is the relative difference in duration to a reference stimulus that has to be exceeded such that the tempos of the two stimuli are perceived differently and was found to be around 5\% relative duration difference~\cite{quene2007just}.
Comparing the two variants to the unmodified case, we see an increased portion of the converted durations inside the \gls{jnd} 
which shows that the approach can indeed perform the desired speaking rate adaptation towards the target speaker.

\begin{table}[h]
    \centering
    \caption{
        Objective evaluation metrics for voice and speaking rate conversion on seen speakers from VCTK.
        Copy-synthesis copies the source speaker vector during conversion, i.e., it reconstructs the source signal.
    }
    \vspace{-3mm}
    \begin{tabular*}{0.45\textwidth}{@{\extracolsep{\fill}}l c c c}
        \toprule[1.5pt]
        Model & EER [\%] $\downarrow$ & WER [\%] $\downarrow$ \\
        \midrule[1pt]
        Natural & $1.83$ & $7.8$ \\ 
        FVAE (copy-synthesis) & $49.90$ & $11.5$ \\ 
        \midrule
        FVAE (no rate mod.) & $6.81$ & $\mathbf{19.5}$ \\ 
        \quad + Scenario 1 & $\mathbf{6.40}$ & $22.3$ \\ 
        \quad + Scenario 2 & $7.01$ & $22.0$ \\ 
        AutoPST (stage 1) & $8.94$ & $49.0$ \\ 
        AutoPST (stage 2) & $7.42$ & $74.1$ \\ 
        \bottomrule[1.5pt]
    \end{tabular*}
    \label{tab:vc-objective}
    \vspace{-1mm}
\end{table}

We follow the methodology by~\cite{das2020predictions} and perform speaker verification and automatic speech recognition to objectively assess the performance of the voice and joint voice and speaking rate conversion.
For each source-target utterance conversion pair, we score both utterances against their respective speaker models (verification target) and against the other speaker (verification non-target) to get a topline \gls{eer} for natural speech.
For \gls{VC} speech, we replace the target utterances with the conversions, i.e., lower \gls{eer} is better.
We use acoustic and language models from ESPNet~\cite{watanabe2018espnet} pretrained on LibriSpeech to report the \gls{wer}.
We use the open-source implementation~\cite{autopst} to train AutoPST where we slightly adapted the feature extraction to work with the ParallelWaveGAN.
Since AutoPST was trained with an one-hot speaker identity vector, we evaluate on the closed speaker test set.
The results are summarized in ~\Cref{tab:vc-objective}.

The WSOLA algorithm introduces some artifacts which negatively impact the \gls{wer}, which nevertheless is still much lower than for AutoPST.
The \gls{wer} for Scenario 1 is slightly higher despite the same voice conversion and time stretching approach as for Scenario 2 due to the stronger interpolations (see also \Cref{fig:duration-diff-vctk}).
Using an interpolation factor for WSOLA that significantly deviates from 1 increases the amount of artifacts.
The bad \gls{wer} for AutoPST stems from the autoregressive spectrogram generation with the Transformer decoder which can arbitrarily shift the content along the time and suffers from repetitions and {elisions} if not well regularized.
We noticed that AutoPST was not able to reliably predict the utterance stop token in many cases which lead to a lot of abnormal speech patterns in the conversions.
The first stage of AutoPST does not target to modify the tempo so we limited the duration of the conversion to its original duration which prevented the \gls{wer} from rising too high.
Adding the speaking rate conversion gives negligible improvements in \gls{eer} which may stem from an increased utterance duration and therefore more accurate speaker statistics.


\vspace{-2mm}
\subsection{Subjective listening tests}
\vspace{-2mm}
In a subjective evaluation\footnote{Audio examples: \url{go.upb.de/interspeech2022}}, we compared the proposed approaches and AutoPST against the same recordings of the converted target speakers. Participants rated the samples on overall quality, and compared their similarity to the reference sample in the dimensions of voice and speaking rate.
The reference recording was modified by the same vocoder used across models \cite{kanbayashiPWG} to allow a more fine grained comparison between the systems.
The reference recording for similarity was the utterance from which the target speaking rate was estimated to ensure a meaningful comparison of speaking rate.
24 different samples were selected randomly and stratified by gender of source speaker, target speaker, as well as the direction of speaking rate conversion (S2F and F2S). To keep the experiment reasonably short, the samples were split into two lists across the source speaker's gender, with half of the participants listening to each.
26 (13 female, 13 male) participants (native speakers of English) were recruited and paid via Prolific.
The collected Mean Opinion Scores (MOS) and standard deviations can be found in \Cref{tab:vc-subjective}. 
\begin{table}[t]
\centering
    \caption{Subjective MOS for overall quality, voice- and speaking rate conversion. Control denotes the optimal topline of the same recording, mixed in with the conversion samples.}
    \vspace{-3mm}
    \resizebox{0.9\textwidth}{!}{\begin{minipage}{\textwidth}
    \begin{tabular}{lrrr}
        \toprule[1.5pt]
        Model & Quality & Simil. Voice & Simil. Speed \\
        \midrule[1pt]
        Control & 4.24 \scriptsize{$\pm$ 0.99} & 4.85 \scriptsize{$\pm$ 0.51} & 4.76 \scriptsize{$\pm$ 0.67} \\
        \midrule
        FVAE (no rate mod.) & \textbf{2.85} \scriptsize{$\pm$ 1.33} & 3.29 \scriptsize{$\pm$ 1.35} & 3.00 \scriptsize{$\pm$ 1.40} \\
        \quad + Scenario 1 & 2.65 \scriptsize{$\pm$ 1.22} & \textbf{3.35} \scriptsize{$\pm$ 1.33} & \textbf{3.29} \scriptsize{$\pm$ 1.47} \\
        \quad + Scenario 2 & 2.79 \scriptsize{$\pm$ 1.29} & 3.33 \scriptsize{$\pm$ 1.33} & 3.10 \scriptsize{$\pm$ 1.45} \\
        AutoPST (stage 1) & 2.56 \scriptsize{$\pm$ 1.36} & 2.82 \scriptsize{$\pm$ 1.39} & 3.01 \scriptsize{$\pm$ 1.47} \\
        AutoPST (stage 2) & 1.81 \scriptsize{$\pm$ 1.15} & 2.88 \scriptsize{$\pm$ 1.44} & 2.41 \scriptsize{$\pm$ 1.48} \\
        \bottomrule[1.5pt]
    \end{tabular}
    \end{minipage}}
    \label{tab:vc-subjective}
\end{table}

To test the statistical significance of the MOS differences, we computed Linear Mixed Effects Regression models \cite{baayen2008mixed} for each quality dimension, with participants added as random effects.
We found a significant improvement in overall quality of FVAE over AutoPST (stage 1) ($\beta$=0.29$^{*}$). 
The voice dimension also showed a significant improvement of FVAE over AutoPST (stage 1) ($\beta$=0.47$^{**}$). This improvement is also significant for the speaking rate adaptation models Scenario~1 ($\beta$=0.53$^{***}$) and Scenario~2 ($\beta$=0.51$^{***}$). 
In the judgements of perceived speaking rate similarity we found a significant interaction between the \gls{VC} model and the direction of speed conversion, so the analysis was carried out independently for both speed conditions. In the S2F case, there were no significant differences in perceived speaking rate scores between the models.
More notably, in the opposing F2S case we found improvements of Scenario~1 over the baseline FVAE ($\beta$=0.65$^{**}$) and AutoPST (stage 1) ($\beta$=0.71$^{***}$).
As observed in the objective evaluations, the failing duration conversion of AutoPST (stage 2) leads to inferior quality and speaking rate distortions as judged by the participants.
This showed to be significant for all quality dimensions, i.e., overall quality (FVAE: $\beta$=-1.03$^{***}$; Scenario~1: $\beta$=-0.84$^{***}$; Scenario~2: $\beta$=-0.98$^{***}$), voice similarity (FVAE: $\beta$=-0.40$^{*}$; Scenario~1: $\beta$=-0.47$^{**}$; Scenario~2: $\beta$=-0.45$^{**}$) and speaking rate similarity (FVAE:$\beta$=-0.87$^{***}$; Scenario~1: $\beta$=-1.52$^{***}$; Scenario~2: $\beta$=-2.08$^{***}$).


\vspace{-1mm}
\section{Conclusion}
This paper investigated the joint conversion of voice and speaking rate.
For voice conversion, we evaluated the performance of a neural disentanglement system that runs fully unsupervised during training and testing.
Speaking rate adaptation was carried out by linear interpolation where the interpolation factor was estimated as the ratio between phoneme rates of source and target utterance.
Adding the speaking rate conversion helped to improve the similarity to the target speaker while only introducing minor artifacts to quality and intelligibility.
Our proposed approach showed to consistently outperform the reference baseline in all metrics and is applicable to unknown speakers.
Future works may investigate the combined use of phoneme and syllable rates for speaking rate conversion which showed to better correlate with perceived tempo~\cite{pfitzinger2002reducing}.
Also, alternative approaches using neural networks~\cite{morrison2021neural, okamoto2022neural} may replace the WSOLA algorithm to reduce the amount of artifacts.

\vspace{-1mm}
\section{Acknowledgements}
This research has been funded by (Deutsche Forschungsgemeinschaft DFG, German Researech Foundation), projects 446378607 and TRR 318/1 2021 – 438445824. Computational resources were provided by the Paderborn Center for Parallel Computing (PC\textsuperscript{2}).

\bibliographystyle{IEEEtran}

\bibliography{mybib}

\end{document}